\documentclass[10pt,onecolumn]{IEEEtran}  

\usepackage{amsmath,amssymb,latexsym,cite}
\usepackage{pstricks,graphicx,subfigure}

\newtheorem{definition}{Definition}
\newtheorem{proposition}{Proposition}
\newtheorem{corollary}{Corollary}
\newtheorem{lemma}{Lemma}
\newtheorem{theorem}{Theorem}
\newcommand{\beq}{\begin{equation}}
\newcommand{\eeq}{\end{equation}}
\newcommand{\bea}{\begin{eqnarray}}
\newcommand{\eea}{\end{eqnarray}}

\newcommand{\EE}{ \mathcal{E}} 

\newcommand{\diag}{\ensuremath{\operatorname{diag}}}

\long\def\symbolfootnote[#1]#2{\begingroup%
\def\thefootnote{\fnsymbol{footnote}}\footnote[#1]{#2}\endgroup}
\newcommand{\Wmat}{\ensuremath{W}}
\newcommand{\Pmat}{\ensuremath{P}}
\newcommand{\ones}{\ensuremath{\vec{1}}}
\newcommand{\defn}{\ensuremath{: \, = }}
\newcommand{\avec}{\ensuremath{\vec{a}}}

\newcommand{\qvec}{\ensuremath{\vec{q}}}
\newcommand{\yvec}{\ensuremath{\vec{y}}}
\newcommand{\eigtwo}{\ensuremath{\lambda_2}}

\newcommand{\order}{\ensuremath{\mathcal{O}}}
\newcommand{\rejeps}{\ensuremath{\mu}}
\newcommand{\rejdelta}{\ensuremath{\nu}}
\newcommand{\Prob}{\ensuremath{\mathbb{P}}}

\newcommand{\Qnum}{\ensuremath{Q}}
\newcommand{\Exs}{\ensuremath{\mathbb{E}}}
\newcommand{\xave}{\ensuremath{\bar{x}_{\operatorname{ave}}}}

\newcommand{\CommCost}{\ensuremath{\mathcal{C}}}

\newcommand{\npar}{\ensuremath{\alpha}}

\long\def\comment#1{}

\newcommand{\myloc}{\ensuremath{\ell}} 
\newcommand{\edge}{\ensuremath{E}}

\begin{document}
\vspace{3in}
\title{Geographic Gossip: Efficient Averaging for Sensor Networks}

\author{Alexandros~G.~Dimakis,~\IEEEmembership{Student
        Member,~IEEE,} Anand~D.~Sarwate,~\IEEEmembership{Student
        Member,~IEEE,}
        and~Martin~J.~Wainwright,~\IEEEmembership{Member,~IEEE}
        \thanks{Manuscript received xxxx; revised yyyy.  The work of
        Alexandros D.~G.~Dimakis was supported by NSF Grants
        CCR-0219722, CCR-0330514 and DMS-0528488. The work of Anand
        D.~Sarwate was supported in part by the NSF Grant CCF-0347298.
        The work of Martin J.~Wainwright was supported by NSF grants
        DMS-0605165 and CCF-0545862. Some of these results were
        presented at the Fifth International Conference on Information
        Processing in Sensor Networks (IPSN 2006).
        \cite{DimakisSW:ipsn06}.  }
\thanks{The authors are with the Department of Electrical Engineering
and Computer Sciences at the University of California, Berkeley.  MJW
is also with the Department of Statistics at the University of
California, Berkeley.}}

\maketitle
\begin{center}
\vspace*{-2.2in}
\vbox to 2.5in{\footnotesize {\tt
 \begin{tabular}[t]{c}
   To appear in \\ 
   IEEE Transactions on Signal Processing. 
  \end{tabular} \vfil}}
\end{center}

\begin{abstract}
Gossip algorithms for distributed computation are
attractive due to their simplicity, distributed nature, and robustness
in noisy and uncertain environments.  However, using standard gossip
algorithms can lead to a significant waste in energy by repeatedly recirculating redundant information.  For realistic
sensor network model topologies like grids and random geometric
graphs, the inefficiency of gossip schemes is related to the slow
mixing times of random walks on the communication graph.  We propose and analyze an alternative gossiping scheme that exploits geographic
information.  By utilizing geographic routing combined with a simple
resampling method, we demonstrate substantial gains over previously
proposed gossip protocols.  For regular graphs such as the ring or
grid, our algorithm improves standard gossip by factors of $n$ and
$\sqrt{n}$ respectively.  For the more challenging case of random
geometric graphs, our algorithm computes the true average to accuracy
$\epsilon$ using $O(\frac{n^{1.5}}{\sqrt{\log n}} \log \epsilon^{-1})$
radio transmissions, which yields a $\sqrt{\frac{n}{\log n}}$ factor improvement over standard gossip algorithms.  We illustrate these theoretical results with experimental comparisons between our algorithm and standard methods as applied to various classes of random fields.
\end{abstract}


\nocite{SpanosOM:05Kalman,Rabbat:06IPSN,AlonBM:87disseminate,HeBSA:03rangefree,MoskAoyamaS:05gossip,PottieK:00,KarpSSV:00rumor,LangendoenR:03localization}

\section{Introduction}

Consider a network of $n$ sensors, in which each node collects a
measurement in some modality of interest (e.g., temperature, light,
humidity).  In such a setting, it is frequently of interest to solve
the \emph{distributed averaging problem}: namely, to develop a
distributed algorithm by which all nodes can compute the average of
the $n$ sensor measurements.  This problem and its connection to
Markov chain mixing rates has been studied for over thirty
years~\cite{deGroot74,Tsitsiklis84}.  It has been the focus of renewed
interest over the past several years, motivated by various
applications in sensor networks and distributed control systems.
Early work~\cite{deGroot74} studied deterministic protocols, known as
consensus algorithms, in which each node communicates with each of its
neighbors in every round.  More recent
work (e.g. \cite{KempeDG:03gossip,BoydGPS:04cdc}) has focused on
so-called gossip algorithms, a class of randomized algorithms that
solve the averaging problem by computing a sequence of pairwise
averages. In each round, one node is chosen randomly, and it chooses
one of its neighbors randomly.  Both nodes compute the average of
their values and replace their own value with this average.  By
iterating this pairwise averaging process, the estimates of all nodes
converge to the global average under suitable conditions on the graph
topology.

The averaging problem is an archetypal instance of distributed signal
processing, in which the goal is to achieve a global objective (e.g.,
computing the global average of all observations) based on purely
local computations (in this case, message-passing between pairs of
adjacent nodes).  Although distributed averaging itself is a very
specialized problem, effective averaging problems provide a useful
building block for solving more complex problems in distributed signal
processing.  Indeed, any averaging algorithm can be easily converted
into a general algorithm that computes any linear projection of the
sensor measurements, assuming that each sensor knows the corresponding
coefficient of the projection vector.  Recently, such algorithms have
been proposed for various problems of distributed computation in
sensor networks, including distributed filtering, detection,
optimization, and
compression~\cite{SpanosOM:05Kalman,XiaoBL:05ipsn,SaligramaTSP,Rabbat:06IPSN}.

A fundamental issue---and the primary focus of this paper---is how
many iterations it takes for any gossip algorithm to converge to a
sufficiently accurate estimate.  These convergence rates have received
significant attention in recent
work~\cite{KarpSSV:00rumor,KempeDG:03gossip,BoydGPS:04cdc,BoydGPS:05Infocom,ChenPX:05ipsn,MoallemiR:05consensus,MoskAoyamaS:05gossip,AlanyaliSS:06computation}. The
convergence speed of a nearest-neighbor gossip algorithm, known as the
\emph{averaging time}, turns out to be closely linked to 
the \emph{spectral gap} (and hence the mixing time) 
of a Markov matrix defined by a weighted random walk on the
graph. Boyd et al.~\cite{BoydGPS:05Infocom} showed how to optimize the
neighbor selection probabilities for each node so as to find the
fastest-mixing Markov chain on the graph. For certain types of graphs,
including complete graphs, expander graphs and peer-to-peer networks,
such Markov chains are rapidly mixing, so that gossip algorithms
converge very quickly.

Unfortunately, for the graphs corresponding to typical wireless sensor
networks, even an optimized gossip algorithm can result in very high
energy consumption.  For example, a common model for a wireless sensor
network is a random geometric graph~\cite{Penrose:03rgg}, in which all
nodes are placed uniformly at random in an area and can communicate with neighbors within some fixed radius $r > 0$.
With the transmission radius scaling in the standard
way~\cite{Penrose:03rgg} as $r(n)=\Theta(\sqrt{\frac{\log n}{n}})$,
even an optimized gossip algorithm requires $\Theta(n^2)$
transmissions (see Section~\ref{Comparisons}), which is of the same
order as the energy required for every node to flood its value to all
other nodes. This problem is noted by Boyd et
al.~\cite{BoydGPS:05Infocom}: ``In a wireless sensor network, Theorem
6 suggests that for a small radius of transmission, even the fastest
averaging algorithm converges slowly'', and this limitation is
intrinsic to standard gossip algorithms applied to such
graphs. Intuitively, the nodes in a standard gossip protocol are
essentially ``blind,'' and they repeatedly compute pairwise averages
with their one-hop neighbors. Information diffuses slowly
throughout the network---roughly moving distance $\sqrt{k}$ in $k$
iterations---as in a random walk.

Accordingly, the goal of this paper is to develop and analyze
alternative---and ultimately more efficient---methods for solving
distributed averaging problems in wireless networks.  We leverage the
fact that sensor nodes typically know their locations, and
can exploit this knowledge to perform geographic routing.
Localization is itself a well-studied problem
(e.g.,~\cite{LangendoenR:03localization, HeBSA:03rangefree}), since
geographic knowledge is required in numerous applications.  With this
perspective in mind, we propose an algorithm that, like a standard
gossiping protocol, is randomized and distributed, but requires
substantially less communication by exploiting geographic
information. The idea is that instead of exchanging information with
one-hop neighbors, geographic routing can be used to gossip with
random nodes who are far away in the network.  The bulk of our
technical analysis is devoted to showing that the resulting rapid
diffusion of information more than compensates for the extra cost of
this multi-hop routing procedure.

In effect, routing to far away neighbors creates an overlay communication network that is the complete graph, where an edge is assigned a cost equal to the
number of hops on the route between the two nodes.  For graphs with
regular topology, it is relatively straightforward to see how this
additional cost is offset by the benefit of faster convergence time.
Indeed, two such examples, the cycle and the grid, are analyzed in
Section~\ref{Algorithm}, where we show gains of the order $n$ and
$\sqrt{n}$ respectively.  The more surprising result of this paper is
that, by using a simple resampling technique, this type of benefit
extends to random geometric graphs---a class of networks with
irregular topology that are commonly used as a model of sensor
networks formed by random deployments.

The remainder of this paper is organized as follows.  In
Section~\ref{Algorithm}, we provide a precise statement of the
distributed averaging problem, describe our algorithm, state our main
results on its performance, and compare them to previous results in
the literature.  In Section~\ref{SecRegular}, we analyze the
performance of our algorithms on two simple regular network
topologies, the cycle and the grid.  Section~\ref{SecRandomGeo}
provides the proofs of our result for the random geometric graph
model.  In Section~\ref{simulations}, we provide a number of
experimental results that illustrate and complement our theoretical
analysis.

\section{Problem formulation and main results}
\label{Algorithm}

In this section, we first formulate the distributed averaging problem
in sensor networks and then describe our algorithm and main
analytical results.  We conclude with an overview and comparison to
related work.

\subsection{Problem statement}
\label{SecProbState}

We begin by formulating the problem of distributed averaging and specifying the technical details of our time and communication models.

\subsubsection{Distributed averaging}

Consider a graph $G$ with vertex set $V = \{1, \ldots, n\}$ and edge
set $E \subset V \times V$.  Suppose that at time $k = 0$, each node
$s \in V$ is given a real-valued number $x_s(0) \in \mathbb{R}$,
representing an observation of some type.  The goal of distributed
averaging is to compute the average $\xave \defn \frac{1}{n}
\sum_{s=1}^n x_s(0)$ at \emph{all nodes} of the graph.  Consensus and
gossip algorithms achieve this goal as follows: at each time slot $k
=0,1,2 \ldots$, each node $s =1, \ldots, n$ maintains an estimate
$x_s(k)$ of the global average.  We use $x(k)$ to denote the
$n$-vector of these estimates; note that that the estimate at different nodes need not agree (i.e., $x_s(k)$ is in general different from $x_t(k)$ for $s \neq t$).  The ultimate goal is to drive the estimate $x(k)$ to the vector of averages $\xave \ones$, where $\ones$ is an $n$-vector of ones.

For the algorithms of interest to us, the quantity $x(k)$ for $k > 0$
is a random vector, since the algorithms are randomized in their
behavior.  Accordingly, we measure the convergence of $x(k)$ to $x(0)$
in the following sense~\cite{KempeDG:03gossip,BoydGPS:05Infocom}:
\begin{definition}
Given $\epsilon > 0$, the $\epsilon$-averaging time is the earliest
time at which the vector $x(k)$ is $\epsilon$ close to the normalized
true average with probability greater than $1 - \epsilon$: 
\begin{align}
\label{ave_time_definition} T_{\mathrm{ave}}(n,\epsilon)=\sup_{x(0)} \inf_{k = 0,1,2 \ldots}
\left\{\Prob \left( \frac{ \|x(k)- x_{ave} \ones \|}{\|x(0)\|} \geq
\epsilon \right) \leq \epsilon \right\},
\end{align}
\end{definition}
where $\| \cdot \|$ denotes the $\ell_2$ norm.  Note that this is
essentially measuring a rate of convergence in probability. \\

\vspace*{.025in}

\subsubsection{Asynchronous time model} 

We use the asynchronous time model \cite{BoydGPS:05Infocom}, which is
well-matched to the distributed nature of sensor networks.  In
particular, we assume that each sensor has an independent clock whose
``ticks'' are distributed as a rate $\lambda$ Poisson process.  The
inter-tick times are exponentially distributed, independent across
nodes, and independent across time.
We note that this model can be equivalently formulated in terms of a
single global clock ticking according to a rate $n\lambda$ Poisson
process.  By letting $Z_k$ denote the arrival times for this global
clock, then the individual clocks can be generated from the global
clock by randomly assigning each $Z_k$ to the sensors according to a
uniform distribution.  On average, there are approximately $n$ global
clock ticks per unit of absolute time (an exact analysis can be found
in \cite{BoydGPS:05Infocom}).  However, our analysis is based on
measuring time in terms of the number of ticks of this (virtual)
global clock.  Time is discretized, and the interval $[Z_k, Z_{k+1})$
corresponds to the $k$th timeslot.  We can adjust time units relative
to the communication time so that only one packet exists in the
network in each time slot with high probability.  Note that this
assumption is made only for analytical convenience; in a practical
implementation, several packets might co-exist in the network, but the
associated congestion control issues are beyond the scope of this
work.

\vspace*{.025in}

\subsubsection{Communication cost}
We compare algorithms in terms of the amount of communication
required.  We will assume a fixed communication radius and hence the number of one-hop radio transmissions is proportional to the total energy spent for communication.  More specifically, let $R(k)$ represent the number of one-hop radio transmissions required for a given node to communicate with some other node in the interval $[Z_k, Z_{k+1})$.  In a standard gossip protocol, the quantity $R(k) \equiv R$ is simply a constant, whereas for our protocol, $R(k)$ will be a random variable (with identical distribution for each time slot).  The total communication cost, measured in one-hop transmissions, is given by the random variable
\begin{align}
\label{EqnDefnCommCost} \CommCost(n, \epsilon) = \sum_{k=1}^{T_{\mathrm{ave}}(n,\epsilon)} R(k)~. 
\end{align}
In this paper, we analyze mainly the expected communication cost,
denoted by $\EE(n,\epsilon)$, which is given by
	\begin{align}
	\EE(n,\epsilon) = E[R(k)] T_{\mathrm{ave}}(n,\epsilon)~.
	\end{align}
Our analysis also yields probabilistic upper bounds on the
communication cost $\CommCost(n, \epsilon)$ of the form 
	\begin{align} 
\Prob \Big \{ \CommCost(n, \epsilon) \geq f(n, \epsilon) \Big \} \leq
\frac{\epsilon}{2}~. 
	\end{align}


\subsubsection{Graph topologies}

\begin{figure}[htb]
\begin{center}
\subfigure[Cycle]{\includegraphics[width=0.20\textwidth]{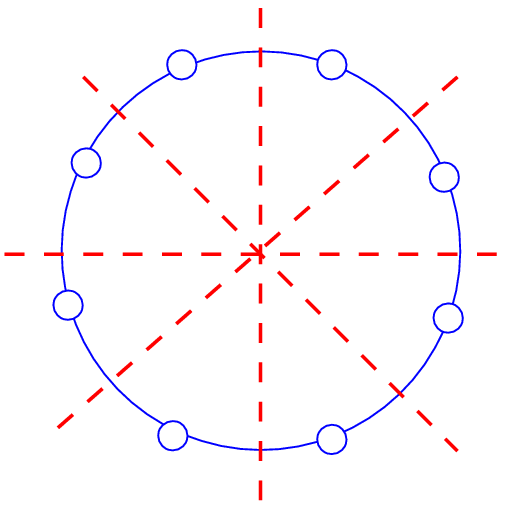}}
\subfigure[Grid]{\includegraphics[width=0.25\textwidth]{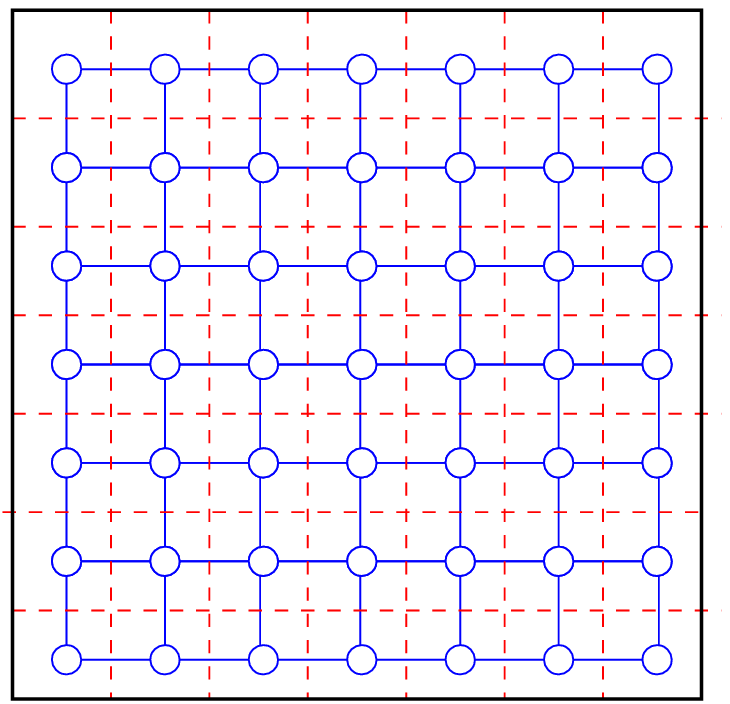}}
\subfigure[Random Geometric Graph]{\includegraphics[width=0.45\textwidth]{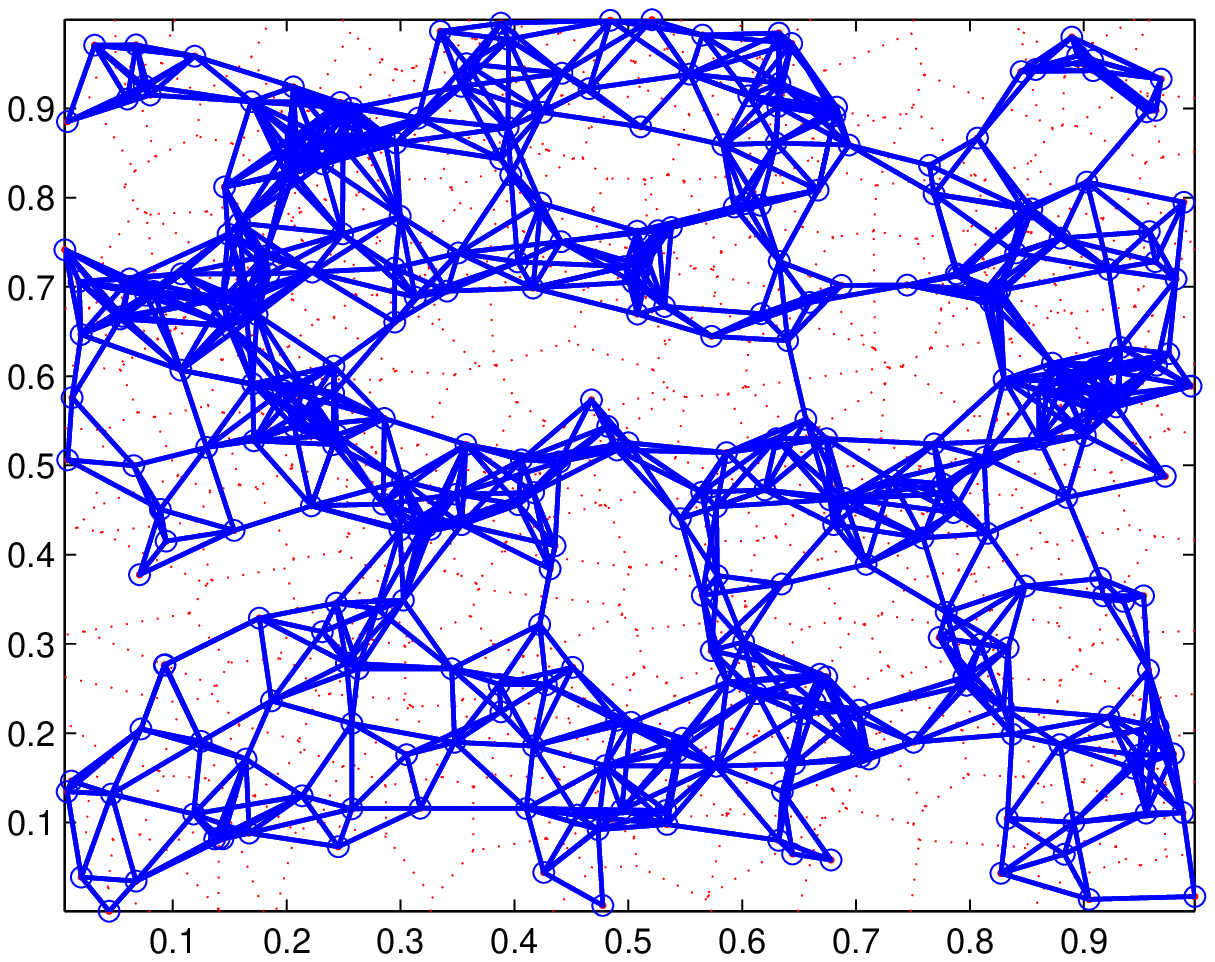}}
\caption{Illustration of a various graphs (nodes as circles and edges
as solid lines) and the associated Voronoi regions associated with
each node (dotted lines).  (a) Cycle graph.  (b) Regular grid.  (c)
Random geometric graph.  }
\label{FigGraphs}
\end{center}
\end{figure}

This paper treats both standard graphs with regular topology,
including the single cycle graph and regular grid as illustrated in
panels (a) and (b) respectively of Figure~\ref{FigGraphs}, and an important subclass of random graphs with irregular topologies, namely those
formed by random geometric graphs~\cite{Penrose:03rgg}.  The random
graph model has been used in previous work on wireless sensor
networks~\cite{GuptaK:00networks, BoydGPS:05Infocom}.  More precisely,
the random geometric graph $G(n,r)$ is formed by choosing $n$ sensor
locations uniformly and independently in the unit square, with any
pair of nodes $s$ and $t$ is connected if and only if their Euclidean
distance is smaller than some transmission radius $r$.  A sample from
this random graph model is illustrated in
Figure~\ref{FigGraphs}(c). It is well known~\cite{Penrose:03rgg,
GuptaK:00networks, ElGamalMPS:04tradeoff} that in order to maintain
connectivity and minimize interference, the transmission radius $r(n)$
should scale like $\Theta(\sqrt{\frac{\log n}{n}})$.  For the purposes
of analysis, we assume that communication within this transmission
radius always succeeds.\footnote{However, we note that our proposed
algorithm remains robust to communication and node failures.}

\subsection{Proposed Algorithm}

The proposed algorithm combines gossip with geographic routing.  The
key assumption is that each node $s$ knows its own geographic location
within some compact subset $C \subset \mathbb{R}^2$, specified as a
Euclidean pair $(x_s,y_s) \in C$.  For the regular grid and random
geometric graphs, we take $C$ to be the unit square $[0,1] \times
[0,1]$, whereas for the single cycle graph we take $C$ to be the unit
circle $S^1$.  In addition, each node can learn the geographic locations of its one-hop neighbors (i.e., vertices $t \in V$ such that $(s,t) \in E)$ using a single transmission per node.


{\bf{Geographic Gossip Algorithm:}}
Suppose the $j$-th clock tick $Z_j$ is assigned to node $s$ at
location $\myloc(s)$.  The following events then happen:
\begin{enumerate}
\item[(1)] Node $s$ activates and chooses a point $y = (y_1, y_2)$
uniformly in the region $C$, referred to as the target location.  Node
$s$ forms the tuple $m_s = (x_s(j), \myloc(s), y)$.
\item[(2)] \label{ggr} Node $s$ sends $m_s$ to its one-hop neighbor $t
\in N(s)$ closest to location $y$.  This operation continues in a
recursive manner: when a successive node $r$ receives a packet $m_s$,
it relays the packet $m_s$ to its one-hop neighbor closest to location
$y$.  Greedy geographic routing terminates when a node receives the
packet and has no one-hop neighbors with distance smaller to the
random target that its own.  Let $v$ be the node closest to location
$y$.
\item[(3)] \label{rej} Node $v$ makes an independent randomized
decision to accept $m_s$. If the packet is accepted, $v$ computes its
new value $x_v(j+1) = \frac{1}{2} (x_v(j) + x_s(j))$ and generates a
message $m_{v} = (x_{v}(j), \ell(v), \ell(s))$, which is sent back to
$s$ via greedy geographic routing.  Node $s$ can then compute its new
value $x_s(j+1) = (x_v(j) + x_s(j))/2$, and the round ends.  If the packet is rejected, then $v$ sends a rejection message to $s$.
\item[(4)] If $v$ rejects the packet from $s$, then $v$ chooses a new
point $y'$ uniformly in the plane and repeats steps
\eqref{ggr}--\eqref{rej} with message $m'_s = (x_{s}(j), \ell(s),
y')$.
\end{enumerate}

\vspace*{.025in}

At a high level, the motivation of the geographic gossip algorithm is
to exploit geographic information (via the greedy routing protocol
described in step (2)) to create a new communication graph $G' = (V,
E')$ as an overlay of the original graph $G = (V, E)$.  Note that the
new communication graph $G'$ has the same vertex set, but an expanded
edge set (i.e., $E' \supset E$).  In fact, for all of the versions of
geographic gossip analyzed in this paper, the extended communication
graph $G'$ is the complete graph, meaning that $(s,t) \in \edge'$ for
all $s \neq t$.  In the standard gossip protocol, each gossip round
takes two radio transmissions.  In the new communication graph $G'$,
certain edges are more costly in terms of one-hop radio transmissions
because of the routing required to carry out the communication.  On
the other hand, the benefit is that the new communication graph $G'$
is dense, so that gossiping converges more quickly.  Our main result
shows that this tradeoff---between the cost of each gossip round and
the total number of rounds---can lead to favorable reductions in the
total number of one-hop radio transmissions.

\subsection{Overview of main results}

The geographic gossip algorithm is a randomized procedure that induces
a probability distribution over the sensor $v$ chosen at each round.
By construction, the probability of choosing sensor $v$ in step (2) of
the geographic gossip algorithm is equal to $a_v$, the area of its
associated Voronoi region.  For certain types of regular graphs, such
as the single cycle and regular grid shown in panels (a) and (b) of
Figure~\ref{FigGraphs}, this distribution over Voronoi regions is
uniform.  In this particularly favorable setting, the ``randomized''
decision of node $v$ in step (3) is simple: it accepts the packet
$m_s$ with probability one.  With this choice, the distribution over
chosen nodes $v$ is guaranteed to be uniform for these regular graphs.
Consequently, it can be shown using known results for mixing on the
complete graph that the averaging time of geographic gossip
$T_{\mathrm{ave}}(n, \epsilon)$ is $\order(n \log \epsilon^{-1})$.
The communication cost given by $\EE(n,\epsilon) = \Exs[R(k)]
T_{\mathrm{ave}}(n,\epsilon)$, where $R(k) \equiv R$ is the number of
single-hop communications required in round $k$ of the protocol.  By
computing the expected value $\Exs[R]$, it can be shown that the
overall communication costs for these regular topologies scale as
$\EE(n, \epsilon) = \Theta(n^2 \log \epsilon^{-1})$ for the single cycle,
and $\EE(n, \epsilon) = \Theta(n^{1.5} \log \epsilon^{-1})$ for the
regular grid.  Thus, as derived in Section~\ref{SecRegular},
geographic gossip yields improvements by factors of $n$ and $\sqrt{n}$
over standard gossip for these regular graphs.

For random geographic graphs, in contrast, the distribution of Voronoi
regions is quite non-uniform.  Consequently, in order to bound the
averaging time $T_{\mathrm{ave}}(n, \epsilon)$, we use in step (3) a
rejection sampling scheme previously proposed by Bash et
al.~\cite{BashBC:04rejsamp} in order to ``temper'' the distribution.
Given the $n$-vector $\avec$ of areas of the sensors' Voronoi regions,
we set a threshold $\tau$. Sensors with cell area smaller than $\tau$
always accept a query, and sensors with cell areas larger than $\tau$
may reject the query with a certain probability.  The rejection
sampling method simultaneously protects against oversampling and
limits the number of undersampled sensors, which allows us to prove
that $T_{\mathrm{ave}}(n, \epsilon) = \order (n \log \epsilon^{-1})$
even for this perturbed distribution.

Of course, nothing comes for free: the rejection sampling scheme
requires a random number $\Qnum$ of queries before a sensor accepts.
Since the queries are independent, $\Qnum$ is a geometric random
variable with parameter equal to the probability of a query being
accepted.  In terms of the number of queries, the total number of
radio transmissions for the $k$th gossip round is $R(k)= \order
\left(\Qnum \cdot G \right)$.  Therefore if $T_{\mathrm{ave}}$ gossip
rounds take place overall, the expected of radio transmissions will be
$\EE(n, \epsilon) = \Exs \left[ \Qnum \cdot G \cdot
T_{\mathrm{ave}}(n,\epsilon) \right]$.  Accordingly, a third key
component of our analysis in Section~\ref{SecRandomGeo} is to show
that the probability of acceptance remains \emph{larger than a
constant}, which allows us to upper bound the expectation of the
geometric random variable $\Qnum$ by a constant.  We also establish an
upper bound on the maximum value of $\Qnum$ over $T_{\mathrm{ave}}$
rounds that holds with probability greater than $1 - \epsilon/2$.

Putting together the pieces yields our main result for random
geometric graphs: the expected cost for computing the average with the
proposed geographic gossip algorithm is
	\begin{align}
	\label{EqnExpMain} 
	\EE(n,\epsilon)= \order \left(\frac{n^{3/2}}{\sqrt{\log n}}
	\log \epsilon^{-1} \right).
	\end{align}
In comparison to previous results on standard gossip for random
graphs~\cite{BoydGPS:05Infocom}, geographic gossip yields a reduction
by a factor of $\sqrt{\frac{n}{\log n}}$ in the number of one-hop
communication rounds.

We note for some classes of graphs, the rejection
sampling may not be necessary, even when the induced distribution is
not uniform, as long as it is reasonably close to uniform.  In
particular, if we have a $\Omega(n^{-1})$ lower bound on the area of a
Voronoi cell for all sensors, then sampling by area is approximately
uniform.  If we can obtain a slightly looser bound on the deviations of the Voronoi areas, alternative techniques may be able to show that our algorithm will not suffer a performance loss without rejection sampling.  However, for geometric random graphs, it is difficult to obtain a good lower bound on the Voronoi cell size, which is our motivation for applying and analyzing the rejection sampling scheme.

\subsection{Related work and comparisons}
\label{Comparisons}

Boyd et al.~\cite{BoydGPS:05Infocom,BoydGPS:04cdc} have analyzed the
performance of standard gossip algorithms.  Their fastest standard
gossip algorithm for the ensemble of random geometric graphs $G(n,r)$
has a $\epsilon$-averaging time~\cite{BoydGPS:05Infocom}
$T_{\mathrm{ave}}(n, \epsilon)= \Theta ( n \frac{\log \epsilon^{-1}
}{r(n)^2})$.  (This quantity is computed in section IV.A of Boyd et
al.~\cite{BoydGPS:05Infocom} but the result is expressed in terms of
absolute time units which needs to be multiplied by $n$ to become
clock ticks).  Consequently, for the standard choice of radius $r(n) =
\Theta (\sqrt{\frac{\log n}{n}})$ ensuring network connectivity, this
averaging time scales as $\Theta (\frac{n^2}{\log n} \log \epsilon^{-1})$.  In standard gossip, each gossip round corresponds to
communication with only one-hop neighbor and hence costs only one
radio transmission which means that the fastest standard gossip
algorithm will have a total cost $\EE(n)= \Theta (\frac{n^2}{\log n}
\log 1/\epsilon)$ radio transmissions.  Therefore, our proposed
algorithm saves a factor of $\sqrt{\frac{n}{\log n}}$ in communication
energy by exploiting geographic information.

A number of recent
papers~\cite{MoallemiR:05consensus,MoskAoyamaS:05gossip,AlanyaliSS:06computation}
have also considered the problem of computing averages in networks.
The consensus propagation algorithm of Moallemi and van
Roy~\cite{MoallemiR:05consensus} is a modified form of belief
propagation that attempts to mitigate the inefficiencies introduced by
the ``random walk'' in gossip algorithms.  For the single cycle graph,
they show improvement by a factor of $\Theta(\frac{n}{\log n})$ over
standard gossip.  Our results for geographic gossip on the single
cycle (see Section~\ref{SecRegular}) show improvement by a factor of
$\Theta(n)$ over standard gossip, and hence a factor $\Theta(\log n)$
over consensus propagation.  It is not yet known how consensus
propagation would behave for the random geometric graphs also
considered in this paper.  Mosk-Aoyama and
Shah~\cite{MoskAoyamaS:05gossip} use an algorithm based on Flajolet
and Martin \cite{FlajoletM:85database} to compute averages, and bound
the averaging time in terms of a ``spreading time'' associated with
the communication graph.  However, they only show the optimality of
their algorithm for a graph consisting of a single cycle, so it is
currently difficult to speculate how it would perform on other regular
graphs or geometric random graphs.  Alanyali et
al.~\cite{AlanyaliSS:06computation} consider the related problem of
computing the average of a network at a \emph{single} node (in
contrast to computing the average in parallel at every node). They
propose a distributed algorithm to solve this problem and show how it
can be related to cover times of random walks on graphs.

\section{Analysis for Regular Networks}
\label{SecRegular}

In this section, we illustrate the benefits of our geographic gossip
algorithm for two simple networks, the ring and the grid, both of
which are regular graphs.  Due to this regularity, the implementation
and analysis of geographic gossip turns out to be especially simple.
More specifically, when these graphs are viewed as contained with the
unit disk (ring graph) or the unit square (grid graph), then the
Voronoi region of each node is equal in area (see
Figure~\ref{FigGraphs}).  Consequently, sampling a \textit{location}
uniformly in the space is equivalent to sampling a \textit{sensor}
uniformly, and thus the overlay graph created by geographic routing
(step (2) of the geographic gossip algorithm) is a complete graph with
uniform edge weights.  In this case, the randomized decision rule in
step~\eqref{rej} is not needed --- the target $v$ always accepts
the message.  For the ring, we show that standard gossip has a
communication cost $\EE(n, \epsilon)$ for $\epsilon$-accuracy that
scales as $\Theta(n^3 \log \epsilon^{-1})$, and that geographic gossip
can improve this to $\order(n^2 \log \epsilon^{-1})$.  For the grid,
we show that standard gossip has communication cost $\Theta(n^2 \log
\epsilon^{-1})$, and geographic gossip can improve this to
$\order(n^{3/2} \log \epsilon^{-1})$.

\subsection{Analysis of single cycle graph}

The ring network consists of a single cycle of $n$ nodes equispaced on
the unit circle (see Figure~\ref{FigGraphs}(a)).  For this simple
network, we have the following result characterizing the improvement
of geographic gossip over standard gossip:
\begin{proposition}
In terms of the communication cost $\EE(n, \epsilon)$ for
$\epsilon$-accuracy, geographic gossip yields a $\Omega(n)$
improvement over standard gossip on the single cycle graph.
\end{proposition}
\begin{proof}
We first compute the communication cost $\EE(n, \epsilon)$ for
standard gossip.  In standard nearest-neighbor gossip, the probability
$p_{ij}$ that nodes $i$ chooses to average with node $j$ is $0$ unless
$|i - j| = 1$, otherwise it is $1/2$.  Therefore the matrix $P =
(p_{ij})$ is a symmetric circulant matrix, generated by the $n$-vector
$(0, 1/2, 0, 0, \ldots, 1/2)$.  Using previous results on standard
gossip~\cite{BoydGPS:05Infocom}, in order to evaluate the performance
of standard gossip, we must find the second eigenvalue $\eigtwo$ of
the matrix $W$ defined by
\begin{align*}
D &= \diag\left( \left\{ \sum_{j=1}^{n} (P_{ij} + P_{ji}) : i = 1,
	2, \ldots, n \right\}\right) \; = \; 2 I \\
W &= I + \frac{1}{2n} D + \frac{1}{2n} (P + P^{T}) \; = \; \left(1 -
	\frac{1}{n} \right) I + \frac{1}{n} P~.
\end{align*}
Note that $W$ is also a circulant matrix, generated by the $n$-vector
$(1 - n^{-1}, (2n)^{-1}, 0, 0, \ldots, (2n)^{-1})$.  Circulant
matrices are diagonalized by the discrete Fourier Transform (DFT)
matrix, so that the eigenvalues can be computed explicitly as
\begin{align*}
\left( 1 - \frac{1}{n} \right) + \frac{1}{n} \cos \frac{k 2 \pi}{n}
	\quad k = 0, 2, \ldots, n-1~.
\end{align*}
Consequently, the second largest eigenvalue is given by
\begin{align*}
\eigtwo(W) &= 1 + \frac{1}{n} \sum_{j=1}^{\infty} (-1)^{2j}
	\frac{1}{(2j)!} \left( \frac{2 \pi}{n} \right)^{(2j)} \; = \;
	1 + \Theta(n^{-3})~.
\end{align*}
Therefore, by a Taylor series expansion, we have $\log \lambda_2(W) =
\Theta(n^{-3})$.  Applying previous results~\cite{BoydGPS:05Infocom}
on standard gossip, we conclude that the $\epsilon$-averaging time of
standard gossip is:
\begin{align*}
T_{\mathrm{ave}}(n, \epsilon) &= \Theta \left(\frac{\log
	\epsilon^{-1}}{\log \eigtwo(W)^{-1}} \right) \; = \;  \Theta \left(
	n^3 \log \epsilon^{-1} \right)
\end{align*}
Since each gossip communication costs us one hop, the average number
of one-hop transmissions for standard gossip on the ring is
\begin{align}
\label{EqnStandardRing}
\EE(n,\epsilon) &=  \Theta \left( n^3 \log \epsilon^{-1} \right).
\end{align}

We now show how geographic gossip reduces the number of one-hop
transmissions.  In geographic gossip for the ring network, a source
node chooses a random location within the unit circle uniformly at random,
which induces a uniform distribution over the nodes in the network
(see Figure~\ref{FigGraphs}(a)).  It then sends a packet to its target
around the ring and they exchange values.  We think of geographic
gossip as running a gossip algorithm on the complete graph with
$p_{ij} = n^{-1}$ for all $i$ and $j$.  For this graph, we have
\begin{align*}
W &= \left( 1 - \frac{1}{n} \right) I + \frac{1}{n^2} \ones \; 
	\ones^T.
\end{align*}
Calculating the second largest eigenvalue yields $\eigtwo(W) = 1 -
\frac{1}{n} + \frac{1}{n^2} = 1 - \Theta(n^{-1})$, so $\log \eigtwo(W)
= \Theta(n^{-1})$, and hence $T_{\mathrm{ave}}(n, \epsilon) = \Theta
\left( n \log \epsilon^{-1} \right)$.  By summing over the pairwise
distances in the graph, we see that the expected number of one-hop
transmissions at any round is bounded by
\begin{align*}
\Exs[R] \; = \; \Exs[R(k)] & \leq \frac{1}{n} \sum_{i=1}^{\lceil
\frac{n}{2} \rceil} (2) = \order(n)~.
\end{align*}
Thus, the expected number of transmissions for geographic gossip is
given by
\begin{align}
\label{EqnGeoRing}
\EE(n,\epsilon) &= T_{\mathrm{ave}}(n, \epsilon) \, \Exs[R] \; = \;
\order \left( n^2 \log \epsilon^{-1} \right)~.
\end{align}
Comparing equations~\eqref{EqnStandardRing} and~\eqref{EqnGeoRing}
yields the claim.
\end{proof}
As demonstrated by this result, for the ring network, using geographic
knowledge and routing improves the energy consumption as measure in
hops by a factor of $n$.  In standard gossip, information from one
node diffuses slowly in a ring, taking almost $n^2$ steps to become
uniformly distributed.  Geographic gossip allows the information from
one node in the network to travel larger distances at the expense of
the routing cost.

\subsection{Analysis of regular grid}

We now turn to geographic gossip on the two dimensional grid defined by a collection of $n$ vertices $s_{ij}$ located at positions $(i/\sqrt{n}, j/\sqrt{n})$ within the unit square $[0,1] \times [0,1]$, as illustrated in Figure~\ref{FigGraphs}(c).
\begin{proposition}
In terms of the communication cost $\EE(n, \epsilon)$ required to
achieve $\epsilon$-accuracy, geographic gossip yields a
$\Omega(\sqrt{n})$ improvement over standard gossip on the regular 2-D
grid.
\end{proposition}
\begin{proof}
The performance of standard gossip on the grid can be calculated using
Corollary 1 from Boyd et al.~\cite{BoydGPS:06it}, which says that the
averaging time is given by $T_{\mathrm{ave}}(n, \epsilon) = \Theta
\left( \frac{n \log \epsilon^{-1}}{1 - \lambda_2(P)} \right)$.  For
standard gossip on the grid, the matrix $P$ is simply the transition
matrix of a random walk on the two-dimensional grid, for which it is
known~\cite{AldousFill} that $(1 - \lambda_2(P))^{-1} = \Theta(n)$.
Consequently, we have $T_{\mathrm{ave}}(n, \epsilon) = \Theta \left(
n^2 \log \epsilon^{-1} \right)$, so that the average number of one-hop
transmissions is
\begin{align}
\label{EqnStandardGrid}
\EE(n,\epsilon) &= \Theta \left( n^2 \log \epsilon^{-1} \right)~.
\end{align}
 
Now let us turn to geographic gossip.  For a regular topology like the
grid, the Voronoi cells are all of equal area, so in step~\eqref{rej}
of the geographic gossip algorithm, the chosen target $v$ simply
accepts with probability one.  Consequently, the number of one-hop
communications per round is simply the route length.  For a regular
2-dimensional grid, routing the message at round $k$ costs $\Exs[R(k)]
= \order(\sqrt{n})$ one-hop transmissions.  As we derived for the ring
network, the geographic gossip algorithm is communicating on an
overlay network that is fully connected, so that the number of rounds
required scales as $T_{\mathrm{ave}}(n, \epsilon) = \order \left( n
\log \epsilon^{-1} \right)$.  Putting the pieces together, we conclude
that the total communication cost for $\epsilon$-accuracy using
geographic gossip scales as
\begin{align}
\label{EqnGeoGrid}
\EE(n,\epsilon) &= \order \left( n^{3/2} \log \epsilon^{-1}
\right)~.
\end{align}
Comparing equations~\eqref{EqnStandardGrid} and~\eqref{EqnGeoGrid}
yields the claim.
\end{proof}

Thus, for the regular grid in 2-dimensions, geographic gossip yields a
factor of $\sqrt{n}$ savings in the convergence time.  The ease of our
analysis in both of the preceding examples---ring and grid
networks---arises from the regularity of the topology, which allowed
us to either write the transition matrix explicitly or use standard
results.  The following section is devoted to analysis of geographic
gossip for random geometric graphs, where we will derive a similar
performance improvement.  For random geometric graphs, in contrast to
the regular topologies considered thus far, we will use a non-trivial
randomized decision rule in step~\eqref{rej} of the gossip algorithm
in order to compensate for irregularities of the graph topology and
areas of Voronoi regions.

\section{Analysis for Random Geometric Graphs}
\label{SecRandomGeo}

We now turn to an analysis of the number of one-hop communications
needed for our algorithm in the case of the random geometric graph
model.  At a high level, our analysis consists of three main steps:
\begin{enumerate}
\item First, we address the number of one-hop transmissions $G$
required to route a packet from node $s$ to the randomly chosen target
$v$ (see step (2) of the geographic gossip algorithm).  We first prove
that when the connectivity radius of the random graphs scales in the
standard way as $r(n) = \Theta (\sqrt{\frac{\log n}{n}})$, greedy
routing always reaches the closest node $v$ to the random target with
\begin{align}
G = \order \left( \sqrt{\frac{n}{\log n }} \right)
\label{g_eq}
\end{align}
one-hop radio transmissions. Note that in practice more sophisticated
geographic routing algorithms (e.g., \cite{KarpK:00gpsr}) can be used
to ensure that the packet approaches the random target when there are
``holes'' in the node coverage. However, greedy geographic routing is
adequate for the problem considered here.
\item As discussed above, when geographic gossip is applied to a graph
with an irregular topology (such as a random geometric graph), it is
necessary to compensate for the irregularity with a non-trivial
accept/reject protocol in step (3) of the algorithm.  Accordingly, our
next step is to bound the expected number of rejections experienced by
a given sensor $s$.
\item The final step is to analyze the number of such gossip rounds
needed for the average to converge to within the target error.  
\end{enumerate}
We take up each of these factors in turn in the subsections to follow.

\subsection{Routing in $\order(1/r(n))$}

We first address how to choose the transmission radius of the sensors
in order to guarantee the network's connectivity and the success of
greedy geographic routing.

\begin{lemma}[Network connectivity]
\label{Connectivity} Let a graph be drawn randomly from the
geometric ensemble $G(n,r)$ defined in Section~\ref{SecProbState}, and
a partition be made of the unit area into squares of side length
$\alpha (n)= \sqrt{ 2 \frac{\log n }{n} }$.  Then the following
statements all hold with high probability:
\begin{enumerate}
\item[(a)] Each square contains at least one node.
\item[(b)] If $r(n)= \sqrt{ 10 \frac{\log n }{n} }$, then each node
can communicate to a node in the four adjacent squares.
\item[(c)] All the nodes in each square are connected with each other.
\end{enumerate}
\end{lemma}

\begin{proof}
The total number of squares of side length $\alpha(n)$ is $M =
\frac{n}{2 \log n}$.  We view these as ``bins'' into which the $n$
sensors are assigned uniformly.  Standard results on this random process \cite{MotwaniR:95randomized,ElGamalMPS:04tradeoff} show that with high
probability $\Theta(M \log M)$ sensors are sufficient to cover all of
the bins, proving (a).


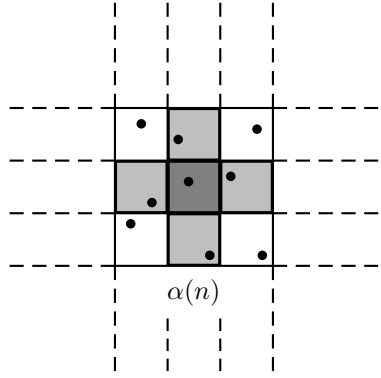
\begin{figure}
\begin{center}
\psset{xunit=0.07cm,yunit=0.07cm,runit=0.07cm}
\begin{pspicture}(0,0)(70,70)

\newgray{medgray}{0.5}
\newgray{lightgray}{0.75}

\psline(20,20)(20,50)
\psline(30,20)(30,50)
\psline(40,20)(40,50)
\psline(50,20)(50,50)

\psline(20,20)(50,20)
\psline(20,30)(50,30)
\psline(20,40)(50,40)
\psline(20,50)(50,50)

\psline[linestyle=dashed](20,20)(0,20)
\psline[linestyle=dashed](20,30)(0,30)
\psline[linestyle=dashed](20,40)(0,40)
\psline[linestyle=dashed](20,50)(0,50)
\psline[linestyle=dashed](50,20)(70,20)
\psline[linestyle=dashed](50,30)(70,30)
\psline[linestyle=dashed](50,40)(70,40)
\psline[linestyle=dashed](50,50)(70,50)

\psline[linestyle=dashed](20,20)(20,0)
\psline[linestyle=dashed](30,10)(30,0)
\psline[linestyle=dashed](40,10)(40,0)
\psline[linestyle=dashed](50,20)(50,0)
\psline[linestyle=dashed](20,50)(20,70)
\psline[linestyle=dashed](30,50)(30,70)
\psline[linestyle=dashed](40,50)(40,70)
\psline[linestyle=dashed](50,50)(50,70)

\psframe[fillstyle=solid,fillcolor=medgray](30,30)(40,40)
\psframe[fillstyle=solid,fillcolor=lightgray](20,30)(30,40)
\psframe[fillstyle=solid,fillcolor=lightgray](40,30)(50,40)
\psframe[fillstyle=solid,fillcolor=lightgray](30,20)(40,30)
\psframe[fillstyle=solid,fillcolor=lightgray](30,40)(40,50)

\rput(35,15){$\alpha(n)$}

\psdots(23,28)(27,32)(25,47)(38,22)(34,36)(32,44)(48,22)(42,37)(47,46)

\end{pspicture}
\caption{Ensuring network connectivity.  Any node in the dark-shaded center square can communicate with its neighbors in the four adjacent lighter-shaded squares if $r(n) = \sqrt{5} \alpha(n)$.  \label{netconnect}}
\end{center}
\end{figure}

Figure \ref{netconnect} shows a simple geometric argument for (b) and
(c).  For $r(n) = \sqrt{5} \alpha(n)$, a sensor at any position in its
square can communicate with all sensor in the four squares adjacent to it.
\end{proof}

\begin{lemma}[Greedy geographic routing]
  \label{routing} Suppose that a node target location is chosen in the
  unit square.  Then greedy geographic routing routes to the node
  closest to the target in $\order( 1/ r(n) )= \order(\sqrt{
  \frac{n}{\log n}})$ steps.
\end{lemma}
\begin{proof}
By Lemma~\ref{Connectivity}(a), every square of side length $\alpha
(n)= \sqrt{ 2 \frac{\log n }{n} } $ is occupied by at least a
node. Therefore, we can perform greedy geographic routing by first
matching the row and then the column of the square which contains the
target, which requires at most $\frac{2}{ r(n)}=\order(\sqrt{
\frac{n}{\log n}})$ hops.  After reaching the square where the target
is contained, Lemma~\ref{Connectivity}(c) guarantees that the subgraph
contained in the square is completely connected.  Therefore, one more
hop suffices to reach the node closest to the target.
\end{proof}

These routing results allow us to bound the cost in hops for an
arbitrary pair of nodes in the network to exchange values.  In the
next section, we analyze a rejection sampling method used to reduce
the nonuniformity of the distribution.

\subsection{Rejection sampling}

As mentioned in the previous section, sampling geographic locations
uniformly induces a nonuniform sampling distribution on the sensors.  Assigning locations to the nearest sensors induces a Voronoi tessellation of the plane, and sensor $v$ is queried with probability proportional to the area $a_v$ of its Voronoi cell.  By judiciously rejecting queries, the sensors with larger Voronoi areas can ensure that they are not oversampled.  We adopt the rejection sampling scheme proposed by Bash et al.~\cite{BashBC:04rejsamp}: when queried, sensor $v$ \emph{accepts} the request with probability
\begin{align}
r_v &= \min\left(\frac{\tau}{a_v}, 1\right)~,
\end{align}
where $\tau$ is a predefined threshold.  Thus sensors with small Voronoi regions always accept, and sensors with large Voronoi regions sometimes reject.

Given $\tau$, the probability $q_v$ that sensor $v$ is sampled can be
written as:
\begin{align}
q_v &= \frac{ \min(\tau, a_v) }{ \sum_{t=1}^{n} \min(\tau, a_t) }
\nonumber \\ 
\label{EqnQvExpress}
&= \frac{ \min(\tau, a_v) }{ |\{t : a_t \ge \tau\}| \cdot \tau +
\sum_{t : a_t < \tau} a_t}~.
\end{align}
\noindent Here the denominator in expression~\eqref{EqnQvExpress} is
the total chance that a query is accepted:
\begin{align}
P_a = \sum_{v=1}^{n} a_v \min\left(\frac{\tau}{a_v}, 1\right) =
|\{v : a_v \ge \tau\}| \tau + \sum_{v : a_v < \tau} a_v~.
\end{align}
\noindent Let $\Qnum$ denote the total number of requests made by
a sensor before one is accepted.


\begin{figure}
\begin{center}
\psset{xunit=0.07cm,yunit=0.07cm,runit=0.07cm}
\begin{pspicture}(-10,-10)(95,55)
\newgray{ltgray}{0.75}

\psline{->}(-5,0)(95,0) \psline{->}(0,-5)(0,55) \psframe(0,0)(5,5)
\psframe(5,0)(10,7) \psframe(10,0)(15,10) \psframe(15,0)(20,15)
\psframe(20,0)(25,15) \psframe(25,0)(30,15) \psframe(30,0)(35,15)
\psframe(35,0)(40,15) \psframe(40,0)(45,15) \psframe(45,0)(50,15)
\psframe(50,0)(55,15) \psframe(55,0)(60,15) \psframe(60,0)(65,15)
\psframe(65,0)(70,15) \psframe(70,0)(75,15) \psframe(75,0)(80,15)
\psframe(80,0)(85,15) \psframe(85,0)(90,15)

\psframe[fillstyle=solid,fillcolor=lightgray](20,15)(25,20)
\psframe[fillstyle=solid,fillcolor=lightgray](25,15)(30,25)
\psframe[fillstyle=solid,fillcolor=lightgray](30,15)(35,27)
\psframe[fillstyle=solid,fillcolor=lightgray](35,15)(40,30)
\psframe[fillstyle=solid,fillcolor=lightgray](40,15)(45,33)
\psframe[fillstyle=solid,fillcolor=lightgray](45,15)(50,37)
\psframe[fillstyle=solid,fillcolor=lightgray](50,15)(55,38)
\psframe[fillstyle=solid,fillcolor=lightgray](55,15)(60,39)
\psframe[fillstyle=solid,fillcolor=lightgray](60,15)(65,42)
\psframe[fillstyle=solid,fillcolor=lightgray](65,15)(70,43)
\psframe[fillstyle=solid,fillcolor=lightgray](70,15)(75,45)
\psframe[fillstyle=solid,fillcolor=lightgray](75,15)(80,47)
\psframe[fillstyle=solid,fillcolor=lightgray](80,15)(85,50)
\psframe[fillstyle=solid,fillcolor=lightgray](85,15)(90,50)

\rput(45,-5){sensor number} \rput{90}(-5,30){cell area}

\psline(-3,15)(3,15) \rput(-7,15){$\tau$}

\end{pspicture}
\caption{Graphical illustration of the rejection sampling
procedure. The total shaded area is the probability of a query being
rejected.  The new sampling distribution is given by the white
histogram, appropriately renormalized.  \label{rejsampfig}}
\end{center}
\end{figure}
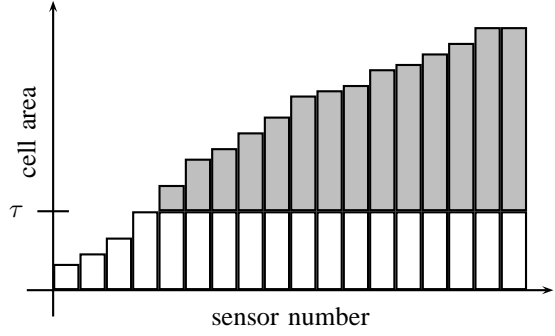

Figure~\ref{rejsampfig} provides a graphical illustration of rejection
sampling on the histogram of Voronoi cell sizes.  Rejection sampling
``slices'' the histogram at $\tau$, and renormalizes the distribution
accordingly.  The total area that is sliced off is equal to $1 -
P_{a}$, the probability that a query is rejected.  Thus, we see that
if $\tau$ is chosen to be too small, then the probability of rejection
becomes very large.  Lemma~\ref{LemGeo} addresses this concern---in
particular, by establishing that the choice $\tau = \Theta(n^{-1})$
suffices to keep the rejection probability suitably bounded away from
$1$, so that the expected number of queries $\Exs[\Qnum]$ remains
finite.  More specifically, we choose $\tau$ such that
\begin{align}
\Prob(a_v \le \tau) = \min\left(\rejdelta, \frac{\rejeps}{1 +
\rejeps}\right)~,
\end{align}
\noindent where the constants $\rejdelta$ and $\rejeps$ control the
undersampling and oversampling respectively.  With this choice of
$\tau$, the results of Bash et al.~\cite{BashBC:04rejsamp} ensure that
no sensor is sampled with probability greater than $(1 + \rejeps)/n$
and no more than $\rejdelta n$ sensors are sampled with probability
less than $1/n$. The following result establishes that the acceptance
probability remains sufficiently large:

\begin{lemma}
\label{LemGeo} 
Ler $0 < c < 1/4$.  For $\tau = c n^{-1}$, we have $\Prob(a_v > \tau) \geq 1 - 4 c$.
\end{lemma}

\begin{figure}
\begin{center}
\psset{xunit=0.07cm,yunit=0.07cm,runit=0.07cm}
\begin{pspicture}(0,0)(60,60)
\newgray{ltgray}{0.75}

\pscircle*(30,30){1} \pscircle[linestyle=dotted](30,30){10}
\pscircle*(18,46){1}
\psline(36,47)(49,34)(37,16)(14,18)(12,29)(36,47)
\psline{->}(30,30)(18,46) \psline{->}(30,30)(30,20)
\rput(35,25){$r$} \pscircle[linestyle=dotted](30,30){20}

\pscircle*(22,4){1} \pscircle*(4,20){1} \pscircle*(55,18){1}
\pscircle*(55,55){1}

\psline(36,47)(36,60) \psline(49,34)(60,36) \psline(37,16)(50,0)
\psline(14,18)(4,0) \psline(12,29)(0,35)

\end{pspicture}
\caption{Inscribing circles in Voronoi cells.  Construction used in
the proof of Lemma~\ref{LemGeo}.  \label{circlebound}}
\end{center}
\end{figure}
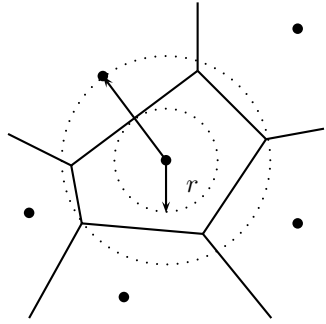

\begin{proof}
We use a simple geometric argument to lower bound $\Prob(a_v > \tau)$.
Consider a node $s$ such that a circle of area $\tau$ it lies
entirely within its Voronoi region, as shown in
Figure~\ref{circlebound}.  Clearly, such nodes are a subset of
those with area larger than $\tau$.  The radius of this circle is $r = \sqrt{\tau/\pi}$.  Note that $r$ is no more than half the distance to the nearest node.
Thus in order to inscribe a circle of radius $\tau$ in the Voronoi
region, all other nodes must lie outside a circle of radius $2r$
around the node. This larger circle has area $4 \tau$, so
\begin{align}
\Prob(a_v > \tau) \ge (1 - 4 \tau)^{n-1} = (1 - 4 c n^{-1})^{n-1}
\geq 1 - 4 c~.
\end{align}
\noindent
Thus, by appropriate choice of $c$, we can make the acceptance
probability arbitrarily close to $1$.
\end{proof}


Our next step is to bound the distance between the new sampling
distribution $\qvec$ (i.e., after tempering by the rejection sampling
procedure), and the uniform distribution $n^{-1}\ones$ over acceptance
regions.  These bounds are used in next section to bound an eigenvalue
of a matrix associated with the gossip algorithm.
\begin{lemma}
\label{rejnorm} For any $\epsilon > 0$, there exists constants
$\rejeps > 0 $ and $\rejdelta > 0$ such that rejection sampling
with parameters $(\rejeps, \rejdelta)$ ensures that
\begin{subequations}
\begin{align}
\label{Eqnlonebound}
  \left\|\qvec - \frac{1}{n} \ones \right\|_1 &< \epsilon, \qquad
  \mbox{and} \\
\label{Eqnltwobound}
  \left\|\qvec - \frac{1}{n} \ones \right\|_2 &< \frac{1}{\sqrt{n}}
  \epsilon~.
\end{align}
\end{subequations}
\end{lemma}
\begin{proof}
Given $\epsilon > 0$, choose $\rejdelta$ and $\rejeps$ such that
$\rejdelta + \rejeps < \epsilon$ and $\rejdelta + \rejeps^2 <
\epsilon^2$. We then expand and bound the error function as
	\begin{align*}
	\sum_{v=1}^{n} \left| q_v - \frac{1}{n} \right| 
		\le \sum_{v: q_v < 1/n} \left| \frac{1}{n} - q_v \right| 
			+ \sum_{v : q_v \ge 1/n} \left| q_v - \frac{1}{n} \right|.
	\end{align*}
\noindent Now we use the properties of rejection sampling from \cite{BashBC:04rejsamp}:
	\begin{align}
		q_v &\le \frac{1 + \rejeps}{n} \qquad \forall v \\
		\left| \left\{ v : q_v < \frac{1}{n} \right\} \right| &\le \rejdelta n~.
	\end{align}
On the set $\{v : q_v \ge 1/n\}$ we use the first bound and on the
set $\{v : q_v < 1/n\}$ we use the second bound:
	\begin{align*}
		\sum_{v=1}^{n} \left| q_v - \frac{1}{n} \right| 
			&\le \left( \rejdelta n \frac{1}{n} 
					+ n \left(\frac{1 + \rejeps}{n} 
						- \frac{1}{n} \right) \right) \\
			&\le \rejdelta + \rejeps~,
	\end{align*}
which is less than $\epsilon$ by our choice of $\rejdelta$ and
$\rejeps$.

Turning now to the bound~\eqref{Eqnltwobound}, we write
\begin{align*}
\left\|\qvec - \frac{1}{n} \ones \right\|_2^2 &= \sum_{v: a_v < \tau} \left| q_v - \frac{1}{n} \right|^2 + \sum_{v : a_v \ge \tau} \left| q_v - \frac{1}{n} \right|^2 \\
&\le \rejdelta n \frac{1}{n^2} + n \left(\frac{\rejeps}{n}\right)^2 \\
&\le \frac{1}{n} (\rejdelta + \rejeps^2) \\ &\le \frac{1}{n}
\epsilon^2~.
\end{align*}
\end{proof}

Finally, we need to bound the expected number of rejections and
the maximum number of rejections in order to bound the expected
number of transmissions and total transmission time.  Recall that
$\Qnum$ is the number of queries that a sensor has to make before
one is accepted, and has a geometric distribution:
	\begin{align}
	\Prob(\Qnum = t) = P_a (1-P_a)^t~.
	\end{align}

\begin{lemma}
\label{exp_rej} For a fixed $(\rejeps, \rejdelta)$, rejection
sampling leads to a constant number of expected rejections.
\end{lemma}
\begin{proof}
The random variable $\Qnum$ is just a geometric random variable
with parameter $P_{a}$, so we can write its mean as:
	\begin{align*}
		\Exs[\Qnum] 
		&= \frac{1}{P_a} \\ 
		&= \frac{1}{|\{v : a_v \ge \tau\}| \tau 
				+ \sum_{v : a_v < \tau} a_v} \\ 
		&\le \frac{1}{(1 - \rejdelta) \tau n} \\
		& = \order(1)~,
\end{align*}
where the final step follows since $\tau = \Theta(n^{-1})$ by
construction.
\end{proof}

\begin{lemma}
\label{max_rej} Let $\{\Qnum_k : k = 1, 2, \ldots K\}$ be a set of
i.i.d. geometric random variables with parameter $P_a$.  For any fixed
pair $(\rejeps, \rejdelta)$, rejection sampling gives
\begin{align}
\max_{1 \le k \le K} \Qnum_k = \order(\log K + \log \epsilon^{-1})
\end{align}
\noindent with probability greater than $1 - \epsilon/2$.
\end{lemma}
\begin{proof}
\newcommand{\maxvar}{\ensuremath{m}}
For any integer $\maxvar \geq 2$, a straightforward computation
yields that
\begin{align*}
\Prob(\Qnum \leq \maxvar-1) = \sum_{t=0}^{\maxvar-1} P_a \,
(1-P_a)^t = 1 - (1-P_a)^\maxvar.
\end{align*}
\noindent By the i.i.d. assumption, we have
	\begin{align*}
		\Prob(\max_k \Qnum_k \leq \maxvar-1) 
		& = \big[1 - (1-P_a)^\maxvar \big]^K \\
		& = \big[1 - \exp (\maxvar \log(1-P_a)) \big]^K~.
\end{align*}

We want to choose $\maxvar = \maxvar(K, \epsilon)$ such that this
probability is greater than or equal to $1 - \epsilon/2$.  First
set $\maxvar = -\rho \frac{\log K}{\log(1-P_a)}$, where $\rho$ is
to be determined.  Then we have
	\begin{align*}
		\Prob(\max_k \Qnum_k \leq \maxvar-1) 
				= \left[1 - 1/K^\rho \right]^K~.
	\end{align*}
We now need to choose $\rho > 1$ such that
	\begin{align*}
	\big[1 - 1/K^\rho \big]^K  &\geq 1 - \epsilon/2~,
	\end{align*}
or equivalently, such that
\begin{align*}
1 - \big[1 - 1/K^\rho \big]^K & \leq \epsilon/2~.
\end{align*}
Without loss of generality, let $K$ be even.  Then by convexity,
we have $(1-y)^K \geq 1 - Ky$.  Applying this with $y = 1/K^\rho$, we
obtain
	\begin{align*}
	1 - \big[1 - 1/K^\rho \big]^K & \leq 1/K^{\rho-1}.
	\end{align*}
Hence we need to choose $\rho \geq \log (2/\epsilon)/\log K + 1$
for the bound to hold.  Thus, if we set
	\begin{align*}
		\maxvar = -\rho \frac{\log K}{\log(1-P_a)} 
			= \order(\log \epsilon^{-1} + \log K)~,
	\end{align*}
then with probability greater than $1-\epsilon/2$, all $K$ rounds of
the protocol use less than $\maxvar$ rounds of rejection.
\end{proof}

\subsection{Averaging with gossip}

As with averaging algorithms based on pairwise updates
\cite{BoydGPS:05Infocom}, the convergence rate of our method is
controlled by the second largest eigenvalue, denoted $\eigtwo(\Wmat)$,
of the matrix
\begin{align*}
\Wmat & \defn I + \frac{1}{2 n} \left[\Pmat + \Pmat^T - D\right]~,
\end{align*}
where $D$ is diagonal with entries $D_i = (\sum_{j=1}^n[\Pmat_{ij}
+ \Pmat_{ji}])$.  The $(i,j)$-th entry of the matrix $\Pmat$ is
the probability that node $i$ exchanges values with node $j$.
Without rejection sampling, $P_{ij} = a_j$, and with rejection
sampling, $P_{ij} = q_j$.  With this notation, we are now equipped
to state and prove the main result of the paper.

\begin{theorem}
\label{Wbound} The geographic gossip protocol with rejection
threshold $\tau = \Theta(n^{-1}$)  has an averaging time
	\begin{align}
		T_{\mathrm{ave}}(n, \epsilon) &= \order \left(n \log \epsilon^{-1} \right)~.
	\end{align}
\end{theorem}

\begin{proof}
To establish this bound, we exploit
Theorem 3 of \cite{BoydGPS:05Infocom}, which states that the
$\epsilon$-averaging time is given by
\begin{align}
T_{\mathrm{ave}}(\epsilon, P) = \Theta \left( \frac{\log
\epsilon^{-1}}{\log \lambda_2(W)^{-1}} \right)~. \label{tave_bound}
\end{align}
Thus, it suffices to prove that $\log \lambda_2(W) = \Omega(1/n)$ in
order to establish the claim.

The probability of any sensor choosing sensor $v$ is just $q_v$, so
that we can write $P$ as the outer product $P = \ones \qvec^T$.
Note that the diagonal matrix $D$ has entries
\begin{align*}
D_i = \sum_{j=1}^n (P_{ij} + P_{ji}) = \sum_{j=1}^{n} q_j +
\sum_{j=1}^{n} q_i = 1 + n q_i~.
\end{align*}
Overall, we can write $W$ in terms of outer products as:
\begin{align}
\label{EqnWdecom} W = \left(I - \diag(\ones + n \qvec)\right) +
\frac{1}{2n} (\ones\  \qvec^T + \qvec\  \ones^T)~.
\end{align}
Note that the matrix $W$ is symmetric and positive semidefinite.

We claim that the second largest eigenvalue $\lambda_2(W) = \order(1 -
c/n)$, for some constant $c$.  By a Taylor series expansion, this
implies that $\log \lambda_2(W) = \Theta(n^{-1})$ as desired. To
simplify matters, we transform the problem to finding the maximum
eigenvalue of an alternative matrix. Since $W$ is doubly stochastic,
Perron-Frobenius theory~\cite{HornJ:87matrix} guarantees that its
largest eigenvalue is one, and has associated eigenvector $v_1 =
n^{-1/2} \ones$.  Consider the matrix $W' = W - \frac{1}{n^2} \ones\
\ones^T$; using equation~\eqref{EqnWdecom}, it can be decomposed as
\begin{align*}
W' = D' + Q',
\end{align*}
where $D' = (I - (2n)^{-1} \diag(\ones + n \qvec))$ is diagonal
and
	\begin{align*}
		Q' = \frac{1}{2n} \ (\ones(\qvec - n^{-1} \ones)^T + (\qvec -
n^{-1} \ones) \ones^T)
	\end{align*}
is symmetric.

Note that by construction, the eigenvalues of $W'$ are simply
	\begin{align*}
		\lambda(W') = \left\{1 - \frac{1}{n}, \lambda_2(W), \ldots,
			\lambda_n(W) \right\}~.
	\end{align*}
On one hand, suppose that $\lambda_1(W') > \lambda_2(W)$; in this
case, then $(1 - \frac{1}{n}) > \lambda_2(W)$ and we are done.
Otherwise, we have
	\begin{align*}
		\lambda_1(W') = \lambda_2(W)~.
	\end{align*}
Note that $W'$ is the sum of two Hermitian matrices -- a diagonal matrix and a symmetric matrix with small entries.  We can therefore apply Weyl's theorem \cite[p.181]{HornJ:87matrix}, to obtain that
	\begin{align*}
		\lambda_1(W') \le \lambda_1(D') + \lambda_1(Q') \le \left(1 -
		\frac{1}{2n}\right) + \lambda_1(Q')~.
	\end{align*}
\noindent It is therefore sufficient to bound $\lambda_1(Q')$. We
do so using the Rayleigh-Ritz theorem \cite[p.176]{HornJ:87matrix},
the Cauchy-Schwartz inequality, and Lemma~\ref{rejnorm} as
follows:
	\begin{align*}
	\lambda_1(Q') 
		&= \max_{\yvec : \|\yvec\|_2 = 1} \yvec^T Q' \yvec \\
		&= \frac{1}{2n} \max_{\yvec : \|\yvec\|_2 = 1} 
				\yvec^T (\ones(\qvec - n^{-1} \ones)^T 
				+ (\qvec - n^{-1} \ones) \ones^T \yvec \\ 
		&= \frac{1}{n} \max_{\yvec : \|\yvec\|_2 = 1} 
				\yvec^T \ones (\qvec - n^{-1} \ones)^T \yvec \\ 
		&\le \frac{1}{n} \max_{\yvec : \|\yvec\|_2 = 1} 
				\|\yvec\|_2 \cdot \|\ones\|_2 
				\cdot \|\qvec - n^{-1} \ones\|_2 
				\cdot \|\yvec\|_2 \\ 
		&\le \frac{1}{n} \left(
				1 \cdot \sqrt{n} \cdot \frac{1}{\sqrt{n}} 
				\epsilon \right) \\ 
		&= \frac{1}{n} \epsilon~.
	\end{align*}
\noindent Overall we have proved the bound
	\begin{align}
	\lambda_1(W') \le \left(1 - \frac{1}{2n}\right) + \frac{1}{n} \epsilon~.
	\end{align}
\noindent We can choose $\epsilon < 1/4$ using Lemma \ref{rejnorm}
to get the desired bound.
\end{proof}

The preceding theorem shows that by using rejection sampling we
can bound the convergence time of the gossip algorithm.  We can
therefore bound the number of radio transmissions required to
estimate the average.
\begin{corollary}
The expected number of radio transmissions required for our gossip
protocol on the geometric random graph $G(n, \sqrt{\frac{\log n}{n}})$
is upper bounded by 
	\begin{align}
	\EE(n,\epsilon) = \order \left(\frac{n^{3/2}}{\sqrt{\log n}} 
			\log \epsilon^{-1} \right)~.
	\end{align}
Moreover, with probability greater than $1 - \epsilon/2$, the maximum
number of radio transmissions is upper bounded
	\begin{align}
		\CommCost(n, \epsilon) =  \order \biggr (\EE(n,\epsilon) 
			\big[ \log n + \log \epsilon^{-1} \big] \biggr)~.
	\end{align}
\end{corollary}
{\bf Remark:} Note that for $\epsilon = n^{-\npar}$ for any $\npar >
0$, our bounds are of the form $\EE(n, 1/n^\npar) = \order (n^{3/2}
\sqrt{\log n})$ and $\CommCost(n, \epsilon) = \order (n^{3/2} \log^{3/2}
n)$.

\begin{proof}
\noindent We just have to put the pieces together. If we assume an
asynchronous protocol, the cost per transmission pair is given by
the product of $\order (\sqrt{n/\log n})$ from routing, $\Exs[\Qnum]$
from rejection sampling, and the averaging time $T_{\mathrm{ave}}$.  From
Lemma~\ref{exp_rej}, $\Exs[\Qnum] = \order (1)$.  Using
equation~\eqref{tave_bound} and Theorem~\ref{Wbound}, we can bound
$\log \lambda_2(W)^{-1}$ by $(1 - \lambda_2(W)) = \order(n^{-1})$.
Thus, the expected number of communications is
	\begin{align}
		\order \left( \sqrt{\frac{n}{\log n}} \Exs[\Qnum] n 
				\log \epsilon^{-1} \right) 
			= \order \left(\frac{n^{3/2}}{\sqrt{\log n}}
				\log \epsilon^{-1} \right)~.
\end{align}
\noindent To upper bound the maximum number of transmissions with
high probability, we note that Lemma~\ref{max_rej} guarantees that
	\begin{align}
		\max_{k=1, \ldots, T_{\mathrm{ave}}} 
			\Qnum_k = \order(\log T_{\mathrm{ave}} 
				+ \log \epsilon^{-1})
	\end{align}
with high probability.  Using Theorem~\ref{Wbound}, we can see that
$\order(\log T_{\mathrm{ave}} + \log \epsilon^{-1}) = \order (\log n + \log
\epsilon^{-1})$. Consequently, with probability greater than $1 -
\epsilon/2$,
\begin{align}
\CommCost(n, \epsilon) = \order \biggr (\EE(n,\epsilon) \big[ \log n +
\log \epsilon^{-1} \big] \biggr)~.
\end{align}
\end{proof}

\section{Simulations}

\label{simulations}

Note that the averaging time is defined in
equation~\eqref{ave_time_definition} is a conservative measure,
obtained by selecting the worst case initial field $x(0)$ for each
algorithm.  Due to this conservative choice, an algorithm is
guaranteed to give (with high probability) an estimated average
that is $\epsilon$ close to the true average for all choices of
the underlying sensor observations.  As we have theoretically
demonstrated, our algorithm is provably superior to standard
gossiping schemes in terms of this metric.  In this section, we
evaluate our geographic gossip algorithm experimentally on
specific fields that are of practical interest.  We construct
three different fields and compare geographic gossip to the
standard gossip algorithm with uniform neighbor selection
probability.  Note that for random geometric graphs, standard
gossiping with uniform neighbor selection has the same scaling
behavior as with optimal neighbor selection
probabilities~\cite{BoydGPS:05Infocom}, which ensures that the
comparison is fair.

\begin{figure}
\begin{center}
\label{experiments}
\includegraphics[width=7cm]{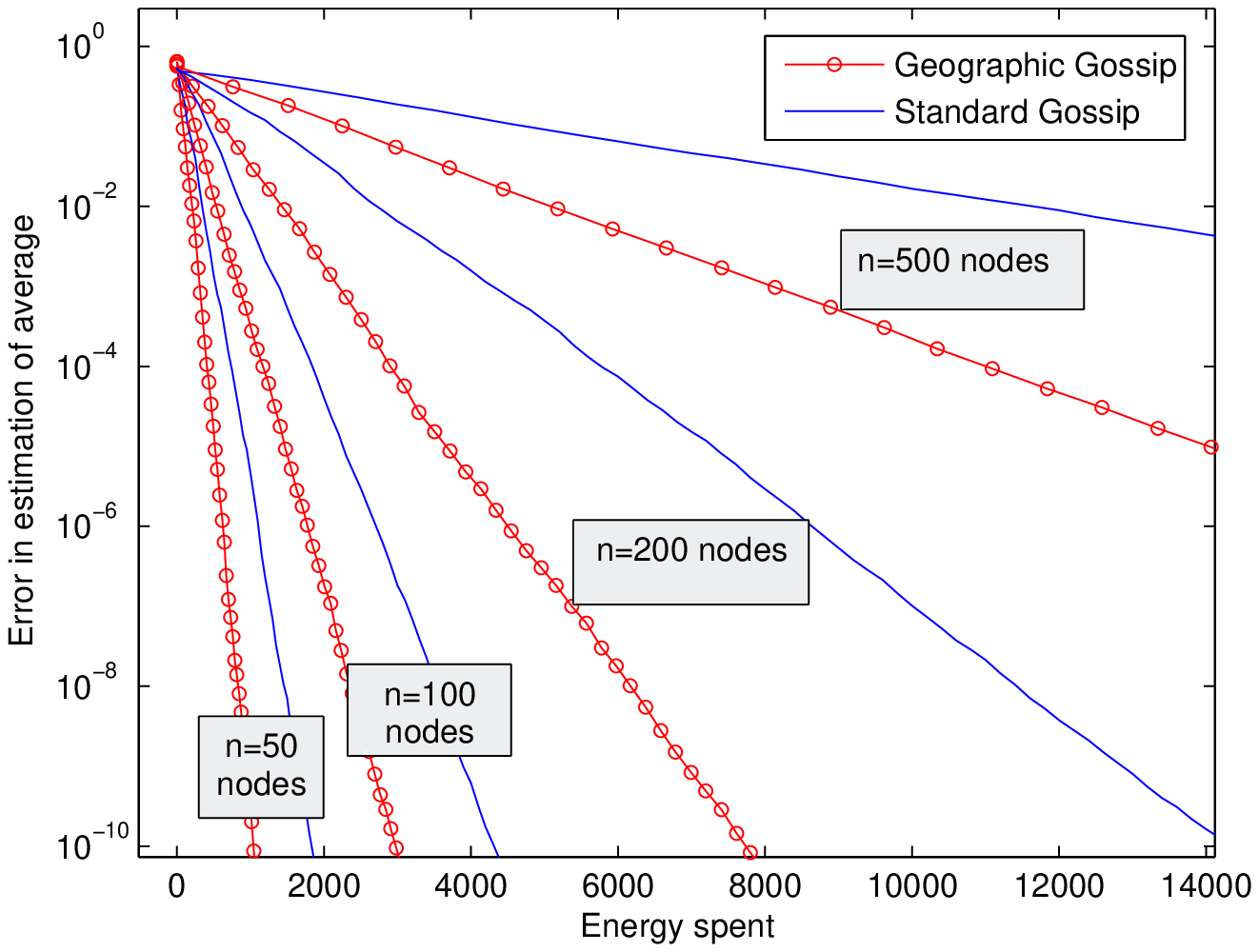}
\includegraphics[width=8cm]{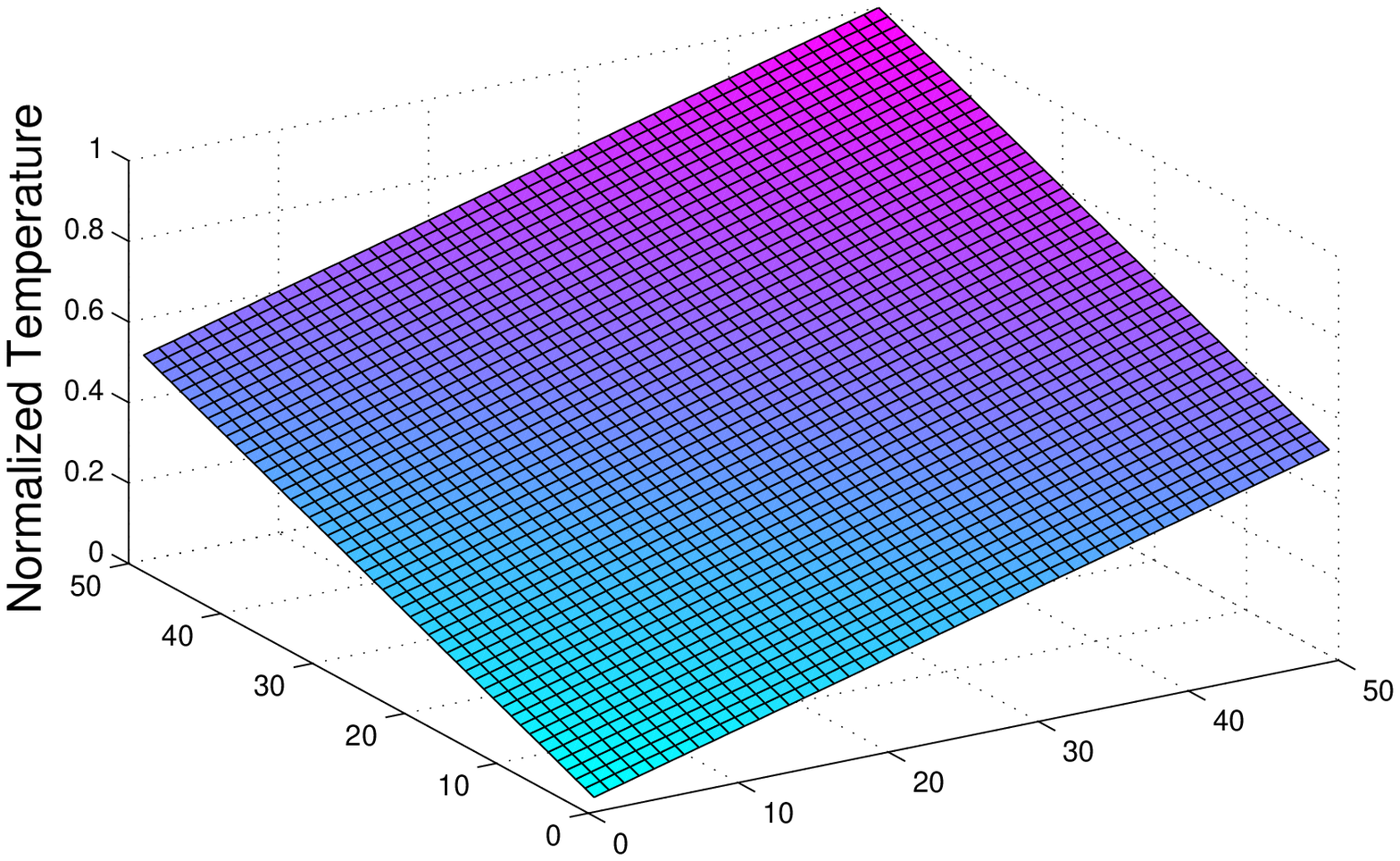}
\caption{Estimation accuracy versus total spent energy for a
linearly varying field.} 
\label{FigA}
\end{center}
\end{figure}

\begin{figure}
\begin{center}
\includegraphics[width=7cm]{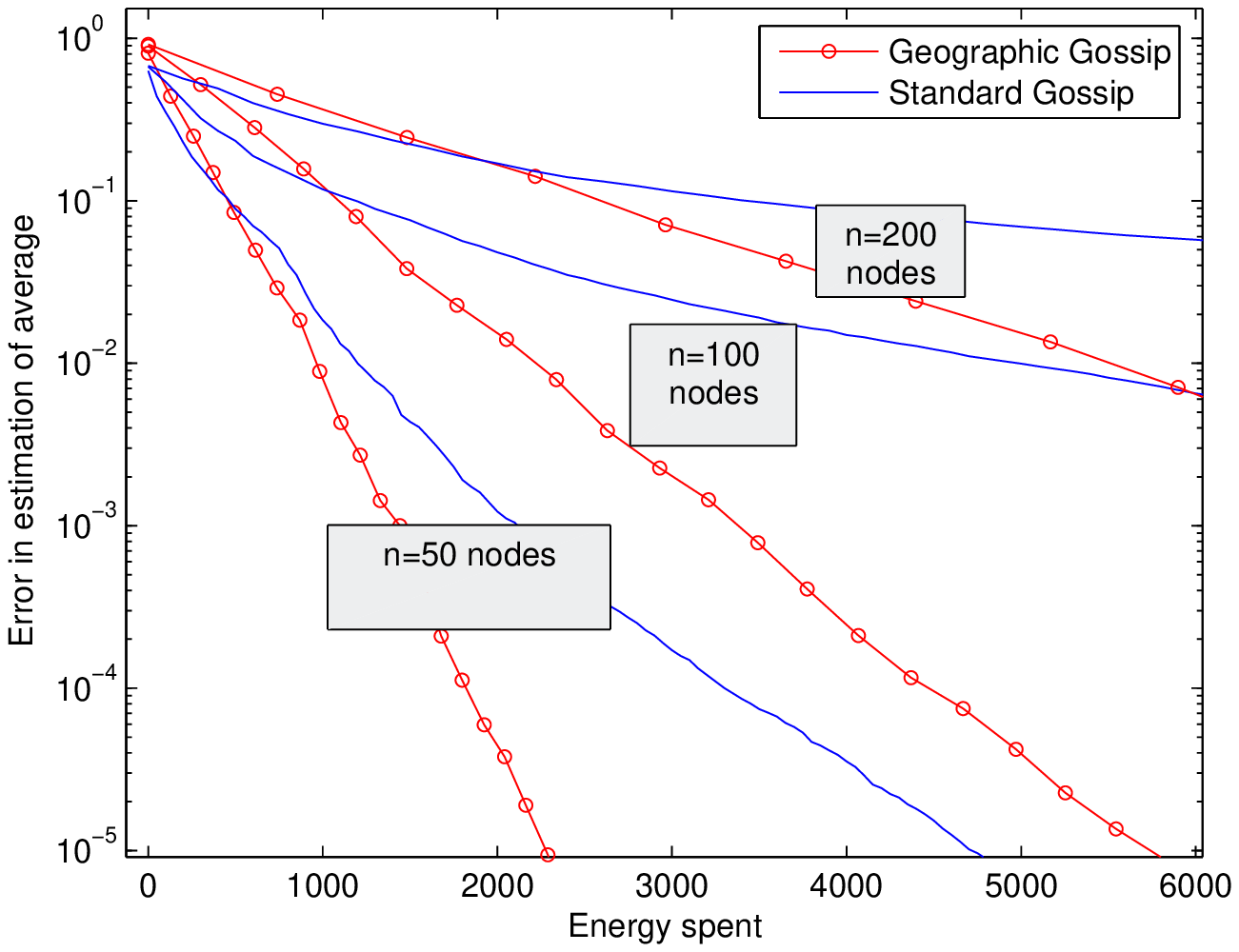}
\includegraphics[width=8cm]{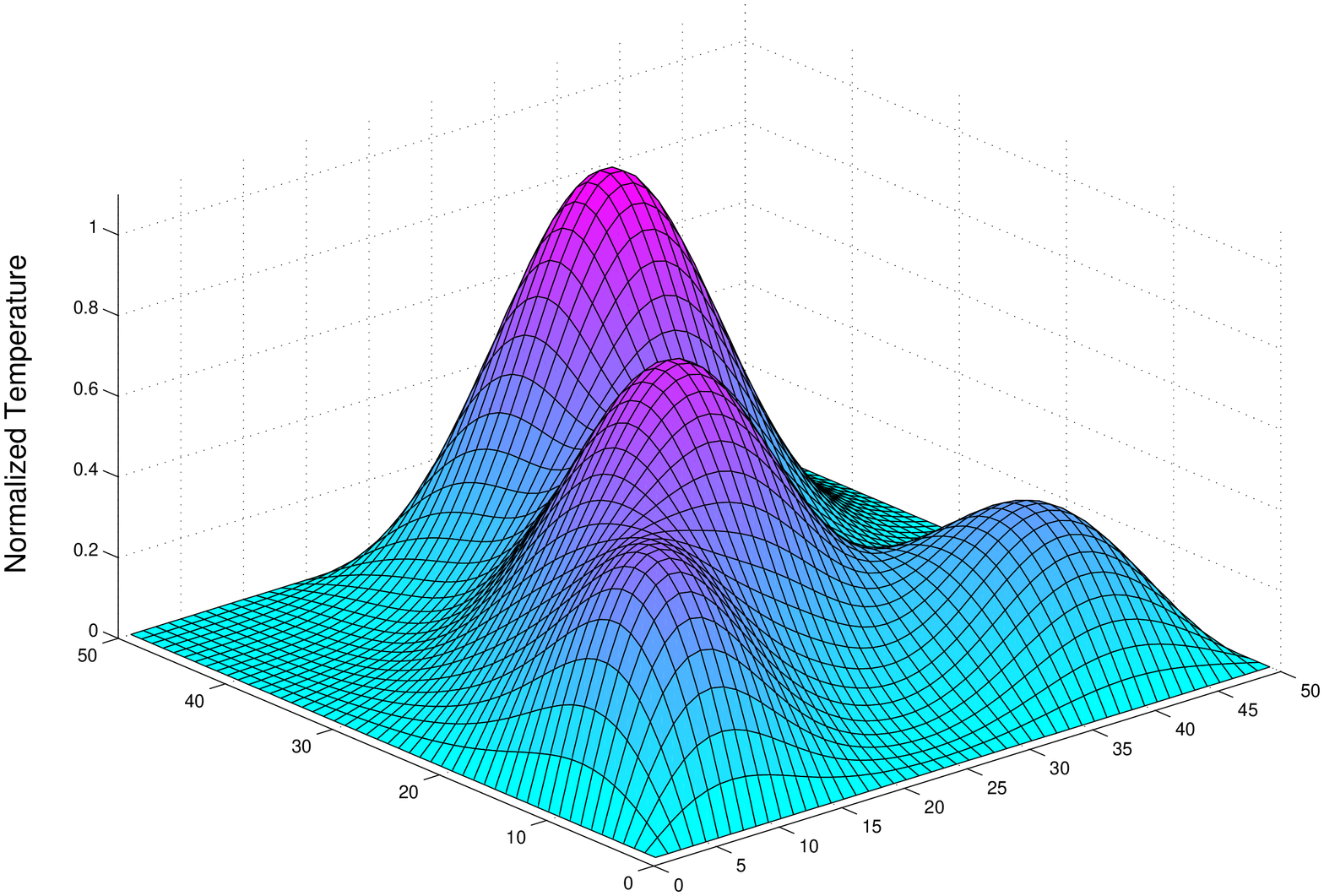}
\caption{Estimation accuracy versus total spent energy for a
smooth field modeling temperature.} \label{FigB}
\end{center}
\end{figure}

\begin{figure}
\begin{center}
\includegraphics[width=7cm]{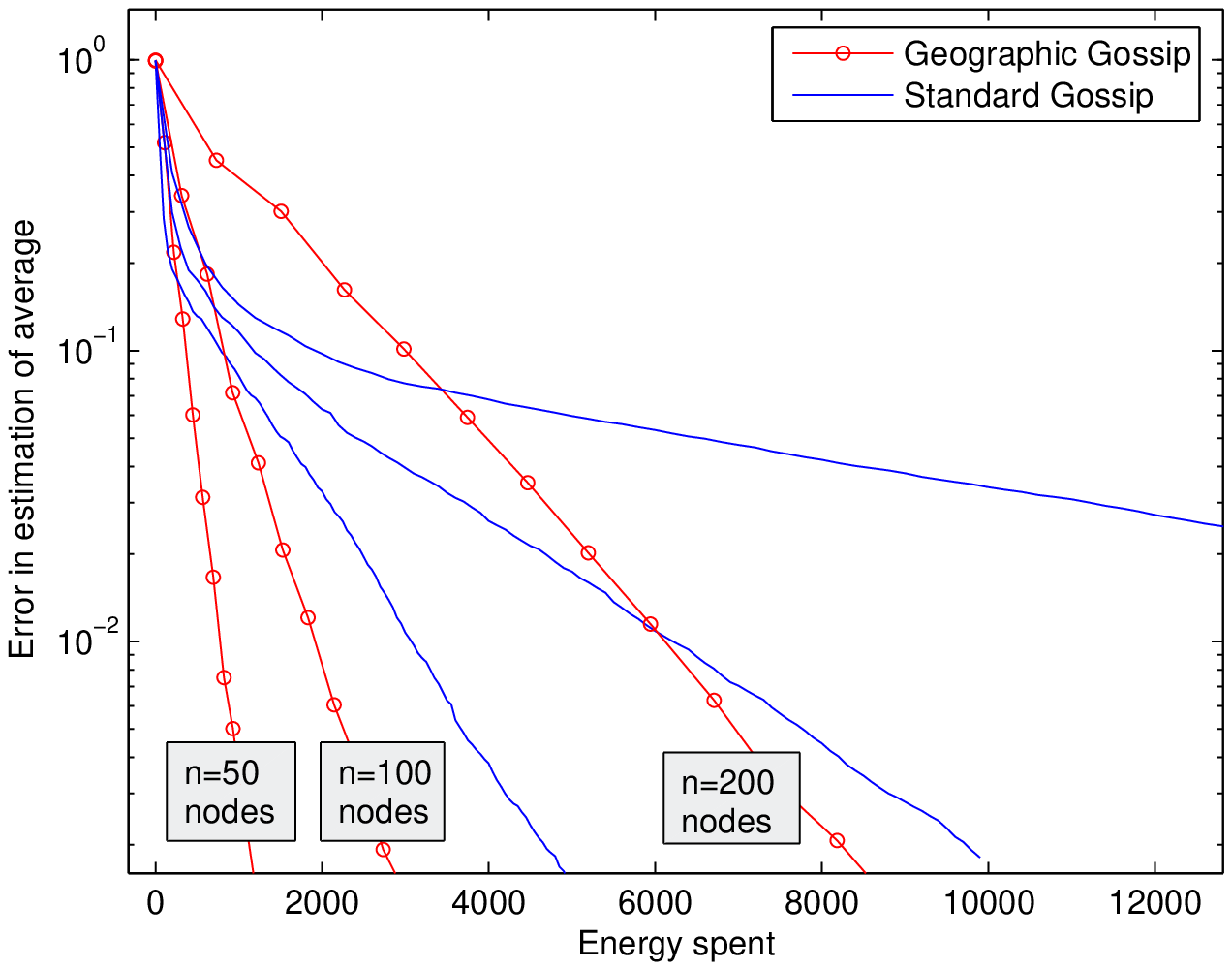}
\includegraphics[width=8cm]{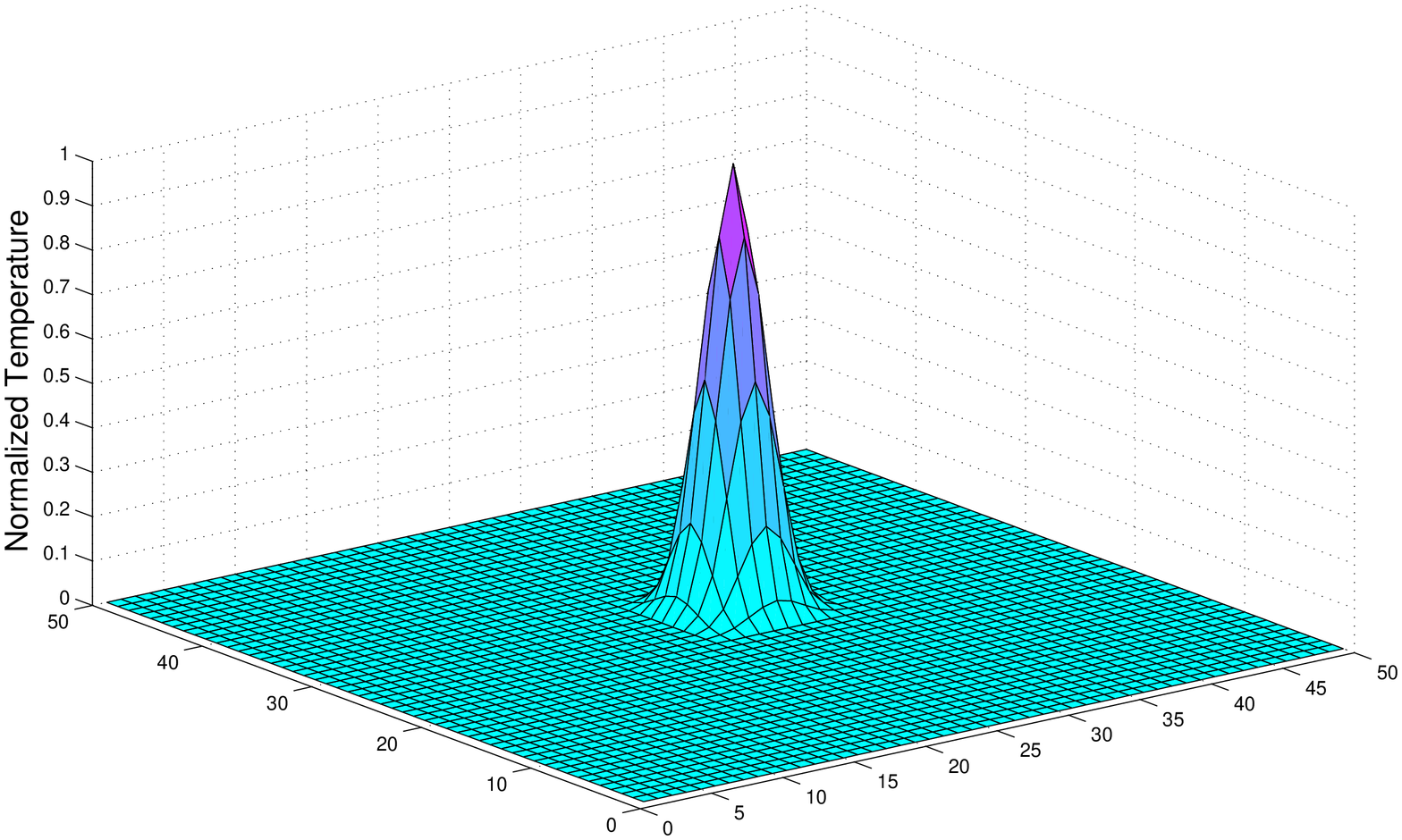}
\caption{Estimation accuracy versus total spent energy for a field
  which is zero everywhere except in a sharp spike.} \label{FigC}
\end{center}
\end{figure}

Figures~\ref{FigA} through~\ref{FigC} illustrate how the cost of each
algorithm behaves for various fields and network sizes. The error in
the average estimation is measured by the normalized $\ell_2$ norm
$\frac{\|x(k)- x_{ave} \ones \|}{\|x(0)\|}$. On the other axis we plot
the total number of radio transmissions required to achieve the given
accuracy.  Figure~\ref{FigA} demonstrates how the estimation error
behaves for a field that varies linearly.  
In Figure~\ref{FigB}, we use a field that is created by
placing temperature sources in the unit square and smooth the
field by a simple process that models temperature diffusion.  Finally,
in Figure~\ref{FigC}, we use a field that is zero everywhere except in
a sharp spike in the center of the field.  For this case, geographic gossip
significantly outperforms standard gossip as the network
size and time increase, except for large estimation tolerances
($\epsilon \approx 10^{-1}$) and small number of rounds.

As would be expected, simple gossip is capable of computing local
averages quite fast. Therefore, when the field is sufficiently smooth,
or when the averages in local node neighborhoods are close to the
global average, simple gossip can generate approximate estimates that
are closer to the true average with a smaller number of
transmissions. For these cases, however, it is arguable that finding
the global average is not of substantial interest in the first place.
In all our simulations, the energy gains obtained by using geographic
gossip were significant and asymptotically increasing for larger
network sizes, corroborating our theoretical results.



\section{Conclusions}

\label{SecDiscussion}

In this paper, we proposed and analyzed a novel message-passing
algorithm for computing averages in networks in a distributed
manner. By exploiting geographic knowledge of the network, our
geographic gossip algorithm computes the averages faster than standard
nearest-neighbor gossip.  Even if the specific type of geographic
routing considered here cannot be performed, similar gossip algorithms
could be developed for any network structure that supports some form
of routing to random nodes.  Thus, our nearest-neighbor gossip can be
understood as a particular case of a more general family of algorithms  in which message-passing occurs on the overlay network supported by random routing.  Other routing protocols may produce different overlay
networks that could be analyzed in a similar manner.

In this paper, we analyzed in detail the case of certain regular
graphs, including the ring and grid networks, as well as the random
geometric graph model, which is commonly used as a model of sensor
networks under random deployments.  Our algorithm can also be applied
to other topologies that realistically model wireless sensor networks,
and should provide gains when (a) the mixing time of a random walk
on the graph is slow (b) efficient routing is possible, and (c)
uniform sampling over space can yield approximately uniform sampling
over sensors.  

Although the current work has focused on the averaging problem, it is
worth noting that many more complicated functions of interest can be
computed using gossip; see the
papers~\cite{MoskAoyamaS:podc06,SpanosOM:05Kalman,RabbatNB:05localization,
XiaoBL:05ipsn} for various examples involving localization, Kalman
filtering and sensor fusion.  However, linear operations (such as
filtering) can be computed using our algorithm by allowing the sensors
to pre-scale their observations by their coefficients in the objective
function.  Our results suggest that geographic gossip may be useful
instead of standard nearest-neighbor gossip to improve energy
consumption in these and other distributed signal processing
applications.

\section{Acknowledgments}

We thank the anonymous reviewers for their careful reading and
constructive criticism that improved the manuscript.

\bibliographystyle{ieeetr}

\bibliography{b_gossipRefs}

\end{document}